\documentclass[showpacs,preprintnumbers,twocolumn]{revtex4}
\usepackage{graphicx}
\usepackage{bm}

\input{tcilatex}

\begin{document}

\preprint{}
\title{Plasmonic nature of van der Waals forces between nanoparticles}
\author{V.V. Klimov}
\email{vklim@sci.lebedev.ru}
\affiliation{P.N. Lebedev Physical Institute, Russian Academy of Sciences, 53 Leninsky
Prospekt, Moscow 119991, Russia}
\author{A. Lambrecht}
\affiliation{Laboratoire Kastler Brossel, ENS, Univ. Paris VI, CNRS, 4, place Jussieu,
Case 74 F-75252 Paris Cedex 05, France}
\keywords{two-sphere cluster, spontaneous emission, plasmon resonance,
nanooptics}
\pacs{}

\begin{abstract}
We propose a new approach to calculate van der Waals forces between
nanoparticles where the van der Waals energy can be reduced to the energy of
elementary surface plasmon oscillations in nanoparticles. The general theory
is applied to describe the interaction between 2 metallic nanoparticles and
between a nanoparticle and a perfectly conducting plane. Our results could
be used to prove experimentally the existence of plasmonic molecules \cite%
{ref1}-\cite{ref2} and to elaborate new control mechanisms for the adherence
of nanoparticles between each other or onto surfaces.
\end{abstract}

\date{\today}
\pacs{42.50.Pq; 73.20.Mf; 85.85.+j; 78.67.Bf}
\maketitle

In recent years, metallic nanoparticles have regained new and vivid
interest. Especially noble-metal nanoparticles are used in biological
sensing applications \cite{ref4}, imaging and as optical materials \cite%
{ref5}. The optical response of nanoparticles is governed by the excitation
of their surface plasmon resonances. Surface plasmons play an important role
in many fields of physics, for example in the enhancement of the
transmission of light through metallic structures \cite{ref6}-\cite{ref8},
or for dispersion forces in 2-dimensional Wigner-like crystals \cite{ref9}.
The properties of surface plasmons and, thus, the optical response of the
nanoparticle depends critically on their material, shape, and size \cite%
{ref10}. For example, optical forces between silver nanoparticle aggregates
are known to be enhanced by exciting conveniently surface plasmon resonances 
\cite{ref11}. Here we investigate a similar phenomenon, that is the
tailoring of van der Waals interaction between nanoparticles via surface
plasmons. This interaction is mediated not by a photon field but by the
electromagnetic vacuum fluctuations, and is particularly interesting for the
design of control mechanisms of the adherence of nanoparticles among each
other or onto a surface.

To describe the van der Waals interactions between a nanoparticle and a
surface, usually a dipole approximation is used \cite{ref25}-\cite{ref26},
which is valid only for large enough distances between the particle and the
surface. Mitchel and Ninham \cite{ref27} suggested to use bispherical
coordinates for the calculation of van der Waals forces between two
spherical particles. Recently van der Waals forces between a spherical
particle and planar substrate have been considered with the help of
expansion of electromagnetic fields over usual spherical harmonics \cite%
{ref28}, while in \cite{ref29}-\cite{ref32} the van der Waals energy between
realistic metallic surfaces was shown to be dominated by surface plasmon
oscillations for small distances.

In the present paper we study the van der Waals energy between two
nanospheres and between a nanosphere and a perfectly conducting plane as the
energy of elementary surface plasmon oscillations in such a system.
Plasmonic oscillations in this system have a very complicated spatial and
spectral structure, which can be described with the paradigm of plasmonic
atoms and plasmonic molecules \cite{ref1}-\cite{ref2}. We will show that
unbound plasmonic atoms and plasmonic molecules give contributions to the
van der Waals energy which are approximately equal in amplitude but opposite
in sign. As a result, the van der Waals energy turns out to be very
sensitive to geometry and materials of the interface and the nanoparticles.
We will consider only short distance limit when all spatial scales are
smaller than the plasma wavelength $\lambda _{pl}=\omega _{pl}$/c, where $%
\omega _{pl}$ is the bulk plasma frequency. We will also assume a local
description of matter and use throughout the paper the dissipationless form
of Drude's dispersion law, $\varepsilon \left( \omega \right) =1-\frac{%
\omega _{pl}^{2}}{\omega ^{2}}$, the plasma model.

To find the plasmon spectrum within the quasistatic approximation we solve
the Laplace equation for potentials $\Delta \Phi =0$ with usual boundary
conditions of continuity for the tangential component of $\mathrm{\mathbf{E}}%
=-\nabla \Phi $ and the normal component of $\mathrm{\mathbf{D}}=\varepsilon 
\mathrm{\mathbf{E}}$ and find resonant values of dielectric permittivity, $%
\varepsilon _{n}$ , n=1,2,3 .., or the corresponding frequencies. It is
obvious that for a specific matter with $\varepsilon =\varepsilon \left(
\omega \right) $ the eigen-oscillations will occur for $\omega _{n}$ such
that $\varepsilon \left( {\omega _{n}}\right) =\varepsilon _{n}$. For the
plasma model, the eigen-frequencies and energy of plasmon fluctuations and
the van der Waals energy will be equal to 
\begin{equation}
\omega _{n}=\frac{\omega _{pl}}{\sqrt{1-\varepsilon _{n}}},U_{vdW}=\sum {%
\frac{\hbar }{2}}\omega _{n}=\frac{\hbar \omega _{pl}}{2}\sum\limits_{n}{%
\frac{1}{\sqrt{1-\varepsilon _{n}}}}  \label{eq2a}
\end{equation}%
The eigen-values of $\varepsilon $ depend on geometrical parameters which
allows to find the van der Waals forces by differentiation over those.
Therefore, the problem is to find universal eigen-values of the dielectric
constant and, then, the frequencies of the plasmonic resonances for specific
materials. We then apply this approach to the van der Waals interaction
between 2 equal spheres and between a sphere and perfectly conducting plane.

\begin{figure}[tbp]
\includegraphics[width=6.2cm] {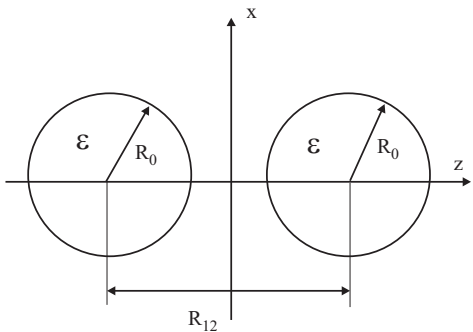}
\caption{Geometry of the van der Waals interaction between 2 spheres.}
\label{fig1}
\end{figure}

Let us consider plasmon oscillations in a 2-sphere cluster with the geometry
shown in Fig.\ref{fig1}. 

\begin{figure}[tbp]
\includegraphics[width=6.2cm] {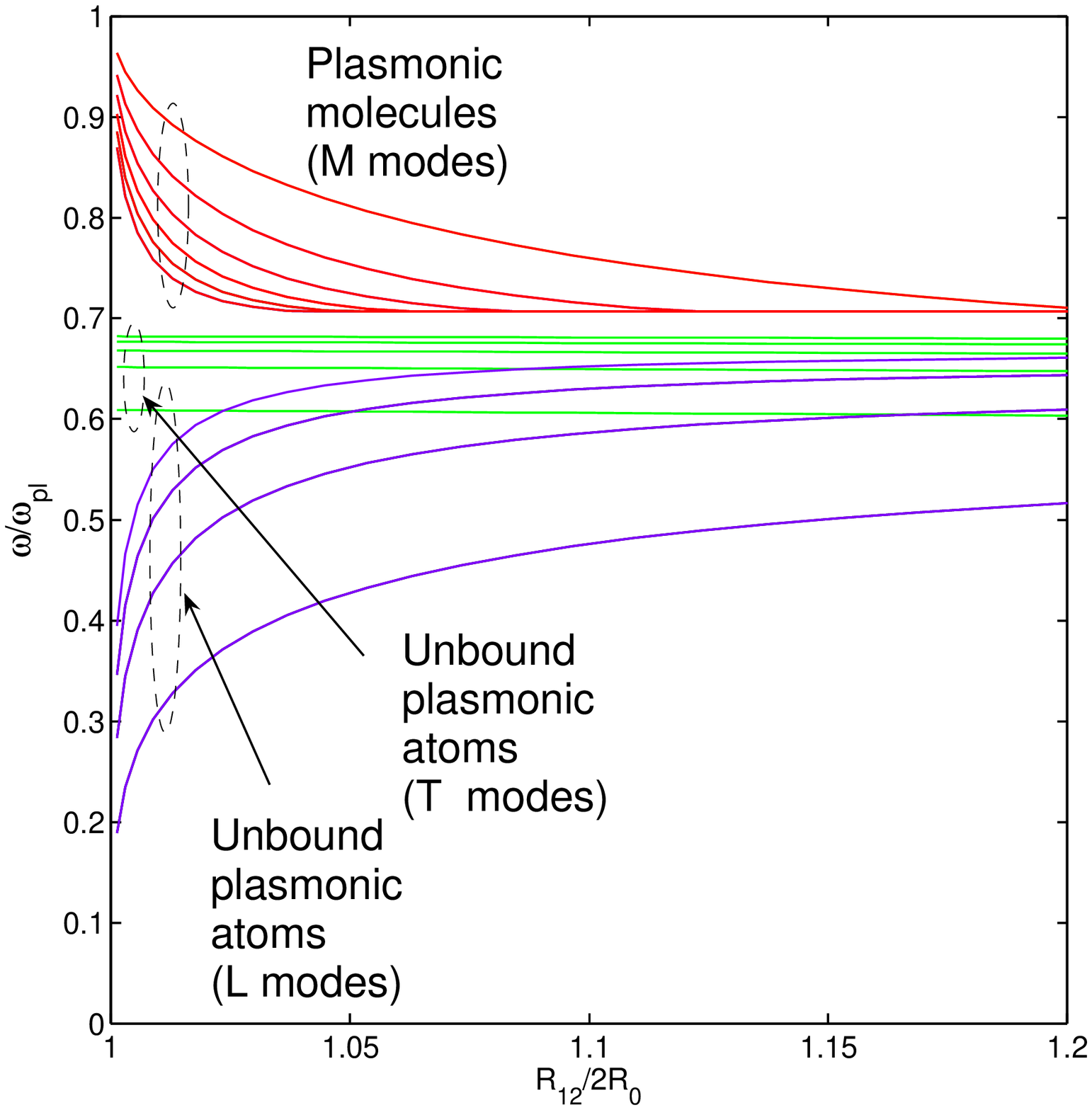}
\caption{Resonant plasmon frequencies as the function of normalized distance
between two identical spheres ($m=0$) with Drude's dispersion law.}
\label{fig2}
\end{figure}

It is natural to use bispherical co-ordinates to calculate full plasmon
spectrum in a two sphere cluster. The result of calculations is shown in
Fig. \ref{fig2} for axially symmetric modes ($m=0$). Other modes have
similar behaviour. 

From this figure one can see that there exist three different types of the
modes: antisymmetrical (with respect to z) L modes and 2 symmetrical M and T
modes. The L and T modes represent unbound states of plasmonic atoms in a
single sphere while M modes represent the bound states of plasmonic atoms,
that is plasmonic molecules \cite{ref1}-\cite{ref2}. These results only
partially agree with previous calculations of plasmon spectrum \cite{ref34}-%
\cite{ref36} which do not predict plasmonic molecules at all. This
difference is not very important in plasmon spectroscopy when a 2 sphere
cluster is excited by uniform (far) optical field with $\omega <\omega _{pl}/%
\sqrt{2}$, because plasmonic molecules do not exist under these conditions.
However, when considering van der Waals interactions ALL the modes should be
taken into account , because the vacuum fluctuations are presented in ALL
the modes. So one should expect substantial difference in van der Waals
forces calculated within different approaches (\cite{ref1}-\cite{ref2} and 
\cite{ref34}-\cite{ref36})

Now consider the properties of different plasmonic modes in more details.
The plasmon frequencies of L modes increase with the sphere separation and
give rise to a van der Waals attraction. On the contrary, the plasmon
frequencies of M and T modes decrease when the distance between the spheres
increases leading to a van der Waals repulsion. Symmetrical T modes are
almost independent of the sphere separation and give no substantial
contribution to the van der Waals energy. Thus, L and M modes give the main
contribution to the van der Waals forces.

\begin{figure}[tbp]
\includegraphics[width=6cm] {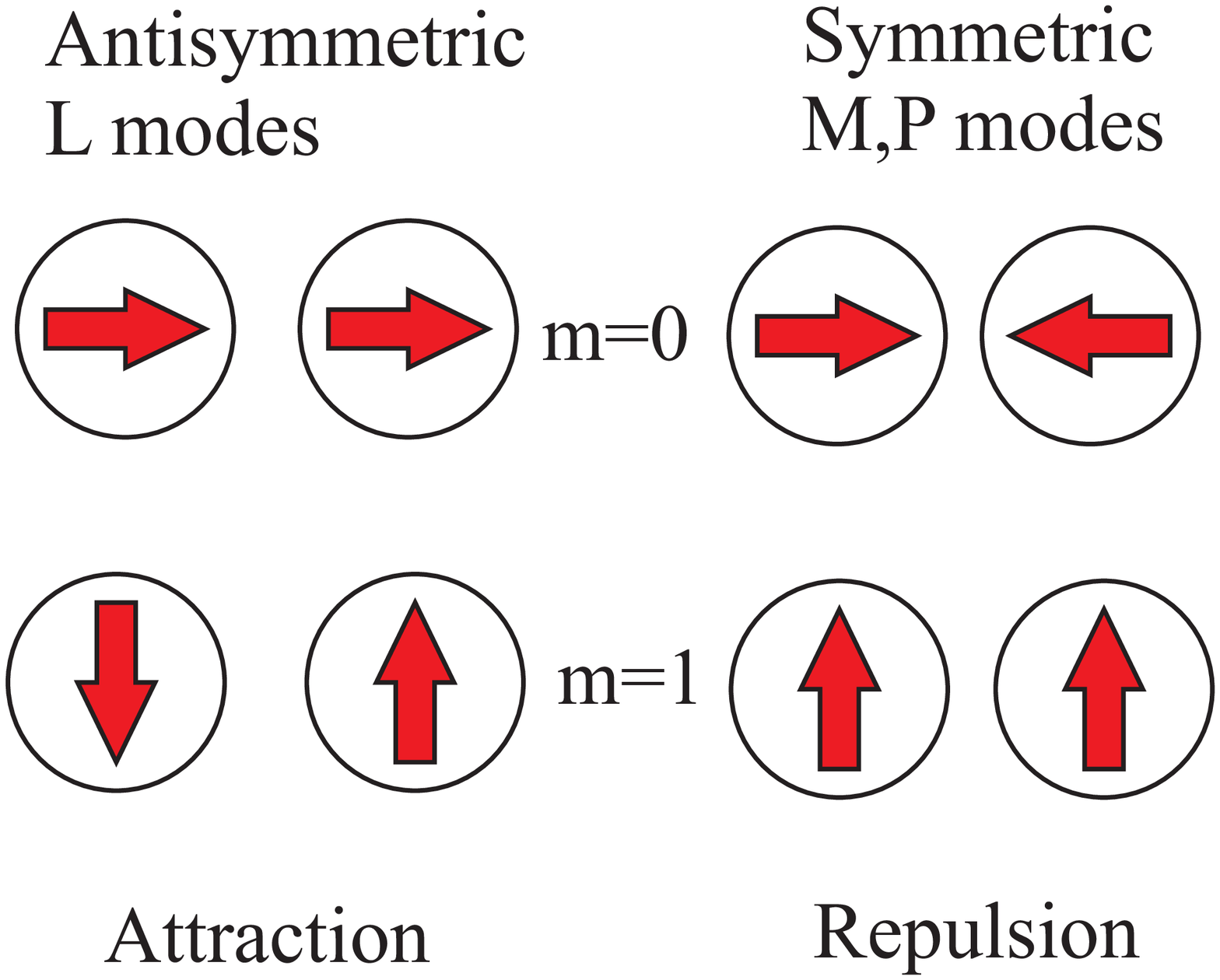}
\caption{Resonant plasmon frequencies as the function of normalized distance
between two identical spheres ($m$=0) with Drude's dispersion law. }
\label{fig3}
\end{figure}

These features of plasmonic modes have a simple physical interpretation
(Fig. \ref{fig3} ). For example, for m=0 (axially symmetrical modes), the
dipole momenta of spheres for any mode is oriented along the z-axis.
However, due to symmetry reasons, the dipole momenta of spheres with excited
L modes are in the same direction, as shown in Fig. \ref{fig3} , while the
dipole momenta of spheres with excited M or T modes have opposite
orientation. Such orientations of dipole momenta result either in attraction
between them, as is the case for the L modes, or in repulsion between them,
such as for M and T modes. For the case $m=1$ (angular dependence $%
cos\varphi $ or $sin\varphi$) we have similar sitiation. 

It is possible to find the following asymptotic expressions for the
eigen-values of the dielectric constant in the case $m\gg 1;L=0,1,2,3,\dots
,m\eta _{0}\sim 1$: 
\begin{equation}
\varepsilon _{m}^{\left( L\right) }=-\frac{\left( {L+m+1}\right) \coth
\left( {\left( {L+m+1/2}\right) \eta _{0}}\right) +\tanh \eta _{0}}{\left( {%
L+m+1}\right) -\tanh \eta _{0}}  \label{eq2}
\end{equation}%
where $\cosh \eta _{0}=R_{12}/\left( 2R_{0}\right) $.

An analogous situation takes place for symmetric M-modes. Asymptotic
expressions for this case have the following form ($m\gg
1,M=0,1,2,\dots,m\eta_{0}\sim 1 $):

\begin{equation}
\varepsilon _{m}^{\left( M\right) }=-\frac{\left( {M+m+1}\right) \tanh
\left( {\left( {M+m+1/2}\right) \eta _{0}}\right) +\tanh \eta _{0}}{\left( {%
M+m+1}\right) -\tanh \eta _{0}}  \label{eq3}
\end{equation}

\begin{figure}[tbp]
\includegraphics[width=7.5cm] {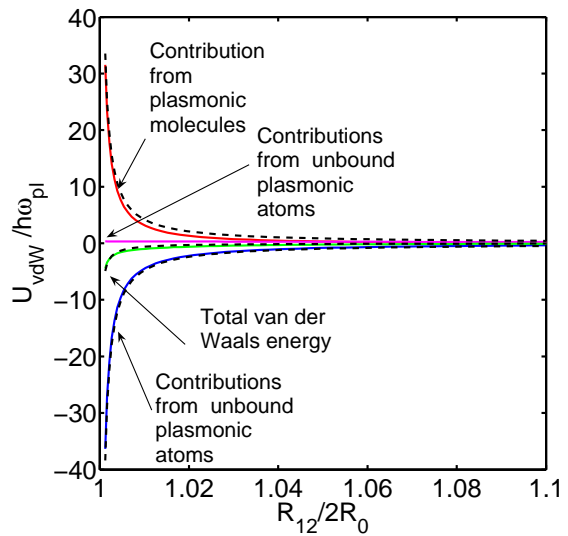}
\caption{The van der Waals energy between 2 equal spheres and different
contributions to it; Dashed lines correspond to asymptotic solutions (%
\protect\ref{eq15}) , (\protect\ref{eq20}) and their sum.}
\label{fig4}
\end{figure}

For a nanosphere and a perfectly conducting plate, plasmonic molecules can
not be excited and only unbound plasmonic atoms (anti-symmetrical L modes)
will contribute to the van der Waals interaction 
\begin{equation}
U_{_{vdW}}^{L}=\frac{\hbar \omega _{pl}}{2}\left[ {\sum\limits_{L=0}^{\infty
}{\frac{1}{\sqrt{1-\varepsilon _{0}^{\left( L\right) }}}}+2\sum%
\limits_{m=1,L=0}^{\infty }{\frac{1}{\sqrt{1-\varepsilon _{m}^{\left(
L\right) }}}}}\right]   \label{eq13}
\end{equation}%
To find the van der Waals force between a sphere and a plate we derive (\ref%
{eq13}) with respect to the distance between the sphere and its mirror
image, that is $R_{12}$ in Fig. \ref{fig1} where the x-axis should be
thought to be replaced by the perfectly conducting plane. To estimate the
van der Waals energy for small distances between the spheres, only large
values of $m$ are important and we can use the asymptotic solution (\ref{eq2}%
) . Substituting (\ref{eq2}) into (\ref{eq13}) and subtracting the vacuum
energy of free space ( $R_{12}\rightarrow \infty ),$ we obtain the following
asymptotic expression 
\begin{equation}
U_{{vdW}}^{L}\approx -0.096\hbar \omega _{pl}\frac{R_{0}}{\Delta }
\label{eq15}
\end{equation}%
where $\Delta $ is the gap width between the sphere and its image. However,
in the region where the sphere is not very close to the plane one has to
calculate plasmonic frequencies of all L modes numerically and then to sum
up the corresponding plasmon frequencies. The result of such summation for $%
m=0...50,L=0...100$ (~ 2*5000 modes) is shown in Fig.\ref{fig4} ( bottom
lines). For the eigenvalue calculation we kept 1500 equations for each $%
m=0..20$ and use asymptotics for $m>20$. The dashed line corresponds to the
asymptotic solution (\ref{eq15}), while the solid line is the result of
numerical calculation. The agreement between asymptote and numerical
solution is very good for small distances. Our results are in agreement with
the proximity theorem but do not agree with results of recent paper \cite%
{ref28}, where a more substantial singularity was predicted.

The van der Waals interaction between two equal nanospheres is more
complicated because now we have to take into account the zero-point energies
of all modes, including plasmonic molecules (M modes) and unbound plasmonic
atoms (L and T modes): 
\begin{equation}
U_{vdW}=U_{vdW}^{M}+U_{vdW}^{L}+U_{vdW}^{T}  \label{eq16}
\end{equation}%
The contribution from L modes was already calculated in Eq.(\ref{eq15}), and
we only have to find the contribution from M and T modes. Beforehand it is
obvious that T modes will give only a small repulsive contribution to van
der Waals forces $U_{\mathrm{vdW}}^{T}\ll U_{vdW}^{M}$. For 
\begin{equation}
U_{\mathrm{vdW}}^{M}=\frac{\hbar \omega _{pl}}{2}\left[ {\sum\limits_{M=0}^{%
\infty }{\frac{1}{\sqrt{1-\varepsilon _{0}^{\left( M\right) }}}}%
+2\sum\limits_{m=1,M=0}^{\infty }{\frac{1}{\sqrt{1-\varepsilon _{m}^{\left(
M\right) }}}}}\right]   \label{eq18}
\end{equation}%
only plasmon frequencies with high values of $m$ will contribute to
asymptotic expressions. We can, thus, use the asymptotic solution (\ref{eq3}%
) to estimate the contribution from M modes to the van der Waals energy.
Substituting (\ref{eq3}) into (\ref{eq18}), we obtain the asymptote 
\begin{equation}
U_{{vdW}}^{M}\approx 0.0838\hbar \omega _{pl}\frac{R_{0}}{\Delta }
\label{eq20}
\end{equation}

The results from numerical summing up of the zero energies of all M and T
modes with $m=0\dots 50,M(T)=0\dots 100$ (~ 4*5000 modes) and the
corresponding asymptotic expressions are shown in Fig.\ref{fig4}, 
which shows a very interesting interplay between repulsive M modes and
attractive L modes. Both sets of modes have the same singularity for close
separation between spheres, but of opposite signs. The amplitudes of these
singularities turn out to be almost equal in value (0.096 and 0.084). As a
result, the full van der Waals energy between spheres remains attractive but
small in comparison with the contributions from both the repulsive plasmonic
molecules (M modes) and attractive unbound plasmonic atoms (L modes). It is
interesting that the small contribution of symmetric plasmonic atoms (T
modes) can not be neglected in this case.

Our approach can be easily generalized to the case of unequal spherical
nanoparticle of the radii $R_{1},R_{2}$. A complete analysis of this general
case will be published elsewhere. Here we restrict ourselves to the case of
closely spaced nanospheres. Again, the main contribution to the van der
Waals forces is due to the modes with large $m,$ and the contributions from
plasmonic molecules and unbound plasmonic atoms can be estimated as 
\begin{equation}
\begin{array}{l}
U_{\mathrm{vdW}}^{M}\approx \hbar \omega _{pl,2}f_{M}\left( {\frac{\omega
_{pl,1}}{\omega _{pl,2}}}\right) \frac{2R_{1}R_{2}}{\Delta \left( {%
R_{1}+R_{2}}\right) } \\ 
U_{\mathrm{vdW}}^{L}\approx \hbar \omega _{pl,2}f_{L}\left( {\frac{\omega
_{pl,1}}{\omega _{pl,2}}}\right) \frac{2R_{1}R_{2}}{\Delta \left( {%
R_{1}+R_{2}}\right) }%
\end{array}
\label{eq22}
\end{equation}%
where $\omega _{pl,1}$ and $\omega _{pl,2}$ are the bulk plasmon frequencies
of the spheres and the functions $f_{M}$ and $f_{L}$ are shown in Fig.\ref%
{fig5}. 
\begin{figure}[tbp]
\includegraphics[width=6cm] {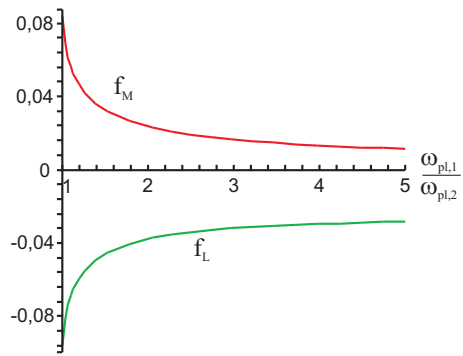}
\caption{Dependence of functions $f_{M}$ and $f_{L}$ (\protect\ref{eq22}) on
ratio $\protect\omega _{pl,1}/\protect\omega _{pl,2}$ of the bulk plasmon
frequencies}
\label{fig5}
\end{figure}
For equal bulk plasmon frequencies (\ref{eq22}) reproduces (\ref{eq15}) and (%
\ref{eq20}). However, for different plasmon frequencies, the contribution
from plasmonic molecules, which give a repulsive force, becomes less
important in comparison to unbound plasmonic atoms, leading to an attractive
force. This fact is in agreement with the case of a nanosphere near a
perfectly conducting plane, where the plasmonic molecules do not contribute
to the van der Waals energy at all.

In the experiment our results could provide additional proof of the
existence of plasmonic molecules by direct measuring of the van der Waals
forces between the nanospheres, clarifying in this way an unsettled
controversy with respect to their existence. Indeed, the calculations made
in \cite{ref34}-\cite{ref36} do not predict symmetrical M modes in plasmon
spectra. As a result, the full van der Waals energy will be described by the
lower curve in Fig.\ref{fig4}. For example, for Al nanoparticles of 50 nm
radius with a gap of about $\Delta $=1nm, van der Waals force within the
plasmon spectrum \cite{ref1}-\cite{ref2} is about 1 nN while it is about 10
nN for the plasmon spectrum predicted in \cite{ref34}-\cite{ref36}. Such a
difference is well measurable with AFM techniques.

On the other hand, a substantial dependence of contributions to the van der
Waals energy from plasmonic molecules and unbound plasmonic atoms on
geometry and materials can have important impact in Nanotechnlogies. The
relative contribution of repelling plasmonic molecules and attracting
unbound plasmon atoms depends critically on the bulk plasmon frequencies of
the nanoparticles. Therefore it would be possible to reduce substantially
the van der Waals forces between different elements and at the same time,
the unwanted phenomena (such as adherence and sticking of nanoparticles), if
one provides the same plasmon frequencies for different elements to enhance
the contribution of repulsing plasmonic molecules.

In conclusion we have calculated the energy of the van der Waals interaction
between 2 closely placed metallic nanospheres as the energy of vacuum
fluctations of ALL plasmonic modes existing in such system. The results
obtained depend crucially on the existence of recently predicted bound
states of plasmons ("plasmonic molecules").This allows us to suggest
experimental measuring of van der Waals forces between 2 nanospheres as the
proof of existence of plasmonic molecules.

\textbf{Acknowledgments}

The authors thank the Russian Basic Research Foundation (V.K., grants {\#}%
07-02-01328, {\#}05-02-19647), Presidium of Russian Academy of Sciences (
V.K., Program "Quantum Macrophysics"), and the University Paris 6 for
partial financial support of this work. A.L. acknowledges partial financial
support by the European Contract STRP-12142 NANOCASE.

\end{document}